# Preparation and characterization of highly thulium- and alumina-doped optical fibers for single-frequency fiber lasers

Honzatko P.[a], Dhar A.[a], Kasik I.[a], Podrazky O.[a], Matejec V.[a], Peterka P.[a], Blanc W.[b], Dussardier B.[b].

[a] Institute of Photonics and Electronic, Chaberska 57, 182 51 Prague, Czech Republic;

[b] Laboratoire de Physique de la Matière Condensée, CNRS UMR6622, Université Nice-Sophia Antipolis, Parc Valrose, 06108 NICE Cedex2, France

e-mail: honzatko@ufe.cz

ABSTRACT

Thulium-doped fibers suitable for core-pumped single-frequency lasers were fabricated by the modified chemical vapor deposition (MCVD) method. Refractive index profile, doping profile and spectral absorption was measured. High doping concentration of thulium ions should be achieved to allow for high absorption of light at a pump wavelength while the thulium ions clustering should be avoided to prevent the cooperative upconversion and quenching processes. The fabricated fibers featured pump absorption up to 70dB/m at a pump wavelength of 1611nm. The single-frequency master oscillator with a resonator composed of a pair of fiber Bragg gratings and a thulium-doped fiber was demonstrated with predominantly single ended operation. We achieved a slope efficiency of 22% and a threshold of 22mW at a lasing wavelength of 1944nm.

## 1. INTRODUCTION

Thulium ions have the widest emission cross-section of any rare-earth ions allowing for operation of continuous and pulse thulium-doped fiber lasers in a wavelength range of 1720-2180nm.1 Numerous applications like remote sensing, material processing, optical coherence tomography can be foreseen.2 While double-clad thulium-doped fibers pumped by multimode high power laser diodes at a wavelength of 793nm are suitable for high power thulium-doped fiber amplifiers, for stable, single-frequency master oscillators, core-pumped fibers with a high concentration of thulium ions are preferable to keep the laser resonator as short as possible to support one mode only. Clustering of the thulium ions should be avoided to prevent the cooperative upconversion and quenching processes.

A single-frequency distributed-feedback thulium-doped silica fiber laser emitting at a wavelength of 1735 nm was demonstrated by Agger et al.3 The laser was pumped from the Ti:sapphire laser at 793 nm and its efficiency was limited by working wavelength to 0.2%. A single-frequency master oscillator (SFMO) made of germanate glass fiber pumped by single-mode laser diode at 805 nm was reported which had a pump threshold of 30 mW, and a slope efficiency of 35%.4 SFMO made of a short piece of a double-clad fiber made of multicomponent glass was demonstrated by Geng et al.5 The multimode pumping together with a short fiber length and a pump wavelength far away from the absorption peak of thulium ions lead to a low slope and a high threshold of this laser. The laser had excellent spectral characteristic with a laser linewidth of 3 kHz.

We present results of preparation and characterization of special silica fibers designed for single frequency master oscillators. The fibers were fabricated by the modified chemical vapor deposition (MCVD) method6 with a thulium concentration in a range 600-2000ppm, a numerical aperture (NA) of 0.15-0.21, and a core diameter of 8-12um. The refractive index profile, doping profile and spectral attenuation were measured for the fabricated samples. The fibers featured the pump absorption up to 70dB/m at a wavelength of 1611nm. The single-frequency master oscillator with a resonator composed of a pair of fiber Bragg gratings and the thulium- doped fiber was demonstrated with predominantly single-ended operation. The active fiber was core-pumped at 1611nm. We achieved a slope efficiency of 22% and a threshold of 22mW at a lasing wavelength of 1944nm.

## 2. FIBER DESIGN, FABRICATION AND CHARACTERIZATION

We fabricated fibers heavily doped by Tm3+ ions. The preforms were prepared by MCVD method. The process started with deposition of porous silica soot layer maintaining suitable deposition temperature inside high quality waveguide silica tube using a standard MCVD setup (Special Gas Controls, GB). Dopant ions namely Al and Tm were introduced in the core-glass matrix employing conventional solution doping technique, where ethanolic solution containing chloride salts of dopant ions (i.e. AlCl3 and TmCl3) was used and soaked for a span of 1 hour. The soaked soot layer was then oxidised, sintered and collapsed to obtain the final perform from which UV-curable acrylated polymer coated fibers of diameter 125±0.5 μm were drawn at a fiber drawing tower. A good solubility of the thulium ions in a glass matrix was achieved by strong doping of the core region with aluminum oxide. The concentration of Al2O3 in the preform core is typically 6-8 mol %.

The geometry of the preforms and fibers is summarized in Table 1. The fibers could principally support a few modes at 2000 nm band. No signs of higher order modes was observed when the input end was carefully spliced with the SMF28 fiber and the fiber was cleaved at the distant end to visualize the far field with a pyroelectric camera (Pyrocam III) at a wavelength of 1945nm. In the laser experiment, the higher order modes do not have sufficient gain to get above the lasing threshold. The mode field diameter (MFD) was estimated based on the divergence measurement using the pyroelectric camera and a linear stage. The 4σ-beam diameter were measured in several slices along the beam path and the MFD was acquired based on the formula

$$MFD \approx \frac{4}{\pi}\theta\lambda \quad (1)$$

where θ is the full divergence angle measured in radians. The MFD measurement method was verified with the SMF28 fiber at 1550 nm, where we measured MFD 10.8 µm that is within the uncertainty limits specified by Corning (10.4±0.8 µm).

The spectral absorption of the fibers is shown in Figure 1. Since the fibers are intended to be pumped in-band, using an L-band communication laser and amplifier, we are interested in the absorption peak above the 1600nm. The position of the absorption peak maximum of the fiber SG1155 is 1657.3nm. All other fibers have narrower absorption peaks with maxima shifted to longer wavelengths then that of SG1155. This indicates lower disorder in the glass matrix. Some characteristics of the absorption peak for the prepared fibers are summarized in the Table 2. For the pumping of the laser we selected a wavelength of 1611 nm. This is a standard communication wavelength in coarse wavelength division multiplexed (CWDM) systems, where CWDM couplers are commercially available. Moreover, the performance of L-band high-power amplifiers is excellent at this wavelength.

The gain of the fibers was measured on short (25-45 cm) fibers at a wavelength of 1945nm. The fiber length was such that almost all the pump power was absorbed. The reabsorption of the signal is small at 1945nm for short fiber, as can be seen from Figure 1. Therefore, we can compare the gain of the fibers even when their lengths are not exactly equal. The gain measurement results are shown in Figure 2. The input power was 2.9mW in all the cases. There are no signs of gain saturation in the measurement range. The gain is generally low, indicating a strong non-radiative loss inherent to thulium-doped fiber amplifiers. It does not preclude an efficient operation in a laser with a high finesse resonator, where the strong signal inside the resonator draws the pump energy. The gain is lower then 1 for low pump powers and the loss is attributed to a pair of splices (0.2-0.8dB per a splice pair depending on the particular sample) and to absorption in the active fiber. The lowest splice loss (≤0.1dB/splice) and the best gain slope was achieved with the fiber SG1151 that has the mode-field diameter well adapted to SMF28 fiber and a better overlap integral between the pump and signal due to the smaller core diameter.

## 3. LASER TEST

The scheme of the single-frequency master oscillator (SFMO) is shown in Figure 3. Until now, we have tested only one of our thulium-doped fibers, namely SG1155. Since this fiber is, according to the gain and pump absorption measurements, one of the worst fibers, even better results can be expected for other fibers. The laser resonator was composed of 23cm of the thulium-doped fiber and a pair of fiber Bragg gratings. The high reflectivity FBG with a bandwidth of 0.3nm and reflectivity 0.99 was placed on the pumping end of the SFMO. The low reflectivity narrow-band FBG with a bandwidth of 0.08nm and reflectivity of 0.95 was used at the output port of the SFMO. As a wavelength selective element for pumping and characterization, standard telecommunication coarse wavelength division multiplexers (CWDM) were used. We used four CWDMs with a pass band centered at 1611nm (Opneti). Their reflection coefficient is about 0.5dB at a lasing wavelength of 1944.5nm. While the first CWDM is not necessary for the operation of the master oscillator, it allows its precise characterization cutting out the L-EDFA amplifier's ASE and at the same time protects the CWDM2 that could be easily burned by the ASE dissipated in the package when operated in the shown orientation. CWDM2 is used to measure the signal ratio between the output port and unused port of the SFMO. A pair of CWDMs 3 and 4 is used to get rid of the unabsorbed pump power.

The Figure 4 shows how the power of the laser output signal increases with the pump power. The signal power is measured at the output as well as pump end of the SFMO. The ratio of these signals is 3.7, proving preferably single end operation of the laser. The slope efficiency (measured on output end only) is 22% with respect to absorbed power, the threshold is 25mW.

## 4. CONCLUSIONS

We have developed an efficient fibers for use in lasers with short resonators. The fibers can be pumped by L-band communication lasers. Their absorption is 42-53 dB/m at a wavelength of 1611 nm allowing to construct lasers with a resonator length of 15-25 cm. Spectral absorption, mode-field diameter, and gain of the fibers was measured. Performance of one of the fibers was tested in a single-frequency master oscillator at 1944 nm. We have achieved a slope efficiency of 20% and a lasing threshold around 20mW. Phase noise characterization as well as further optimization of the master oscillator is under way.

## ACKNOWLEDGEMENT

The work was supported by the Czech Science Foundation project P205-11-1840.

Table 1. Geometry of the preforms and fibers.

| Fiber | $Tm^{3+}$ [ppm] | NA | Preform core [mm] | Preform cladding [mm] | Fiber core [μm] | MFD [μm] |
|---|---|---|---|---|---|---|
| SG1151 | 1140 | 0.20 | 1.06 | 9.3 | 14.2 | 9.2 |
| SG1155 | 1530 | 0.19 | 1.01 | 9.4 | 13.0 | 10.5 |
| SG1172 | 1300 | 0.22 | 1.06 | 9.2 | 14.4 | 11.9 |
| SG1173 | 940 | 0.21 | 0.99 | 9.0 | 14.2 | 12.0 |

Table 2. Characteristics of $Tm^{3+}$ absorption peak corresponding to the transition $^3F_4$-$^3H_6$

| Fiber | Peak [nm] | Width [nm] | Peak absorption [dB/m] | Absorption@1611 nm [dB/m] |
|---|---|---|---|---|
| SG1151 | 1662.4 | 58.8 | 58.3 | 48.4 |
| SG1155 | 1657.6 | 65.7 | 49.0 | 42.3 |
| SG1172 | 1661.3 | 55.9 | 64.2 | 53.6 |
| SG1173 | 1661.5 | 58.4 | 58.1 | 48.6 |

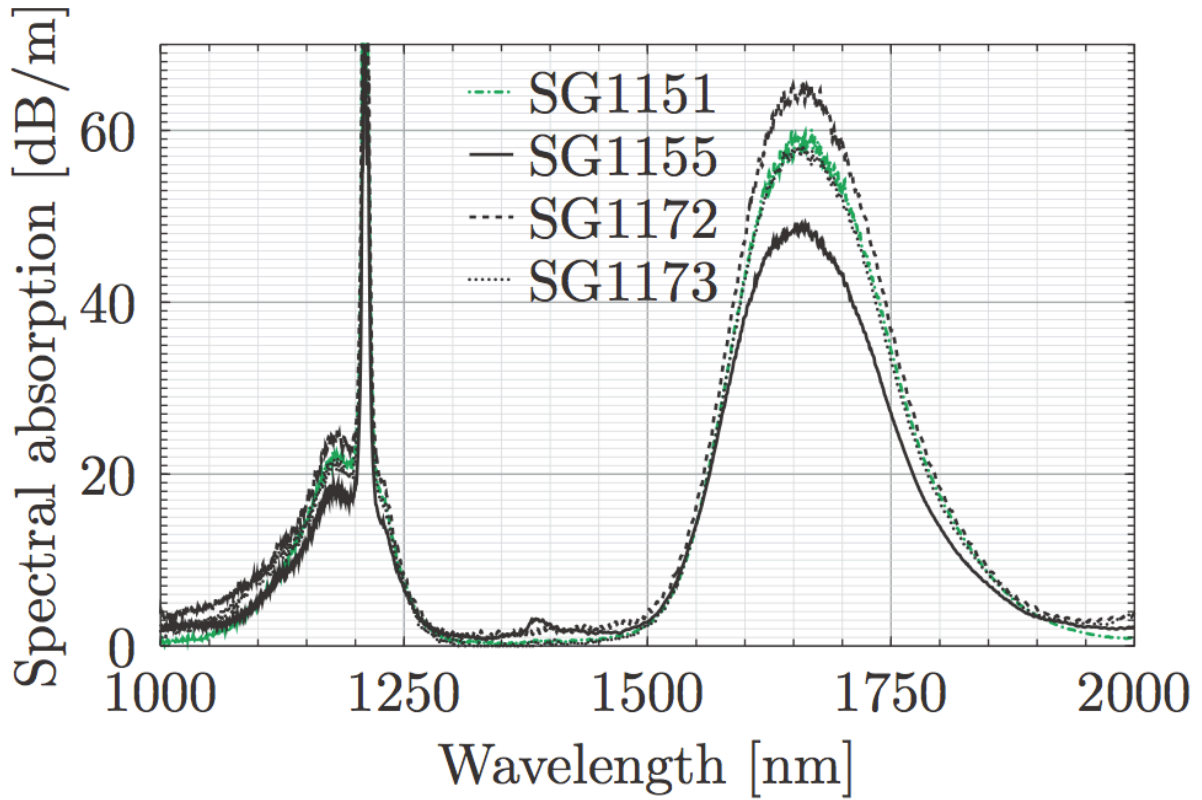

Figure 1. Spectral absorption of the fabricated fibers.

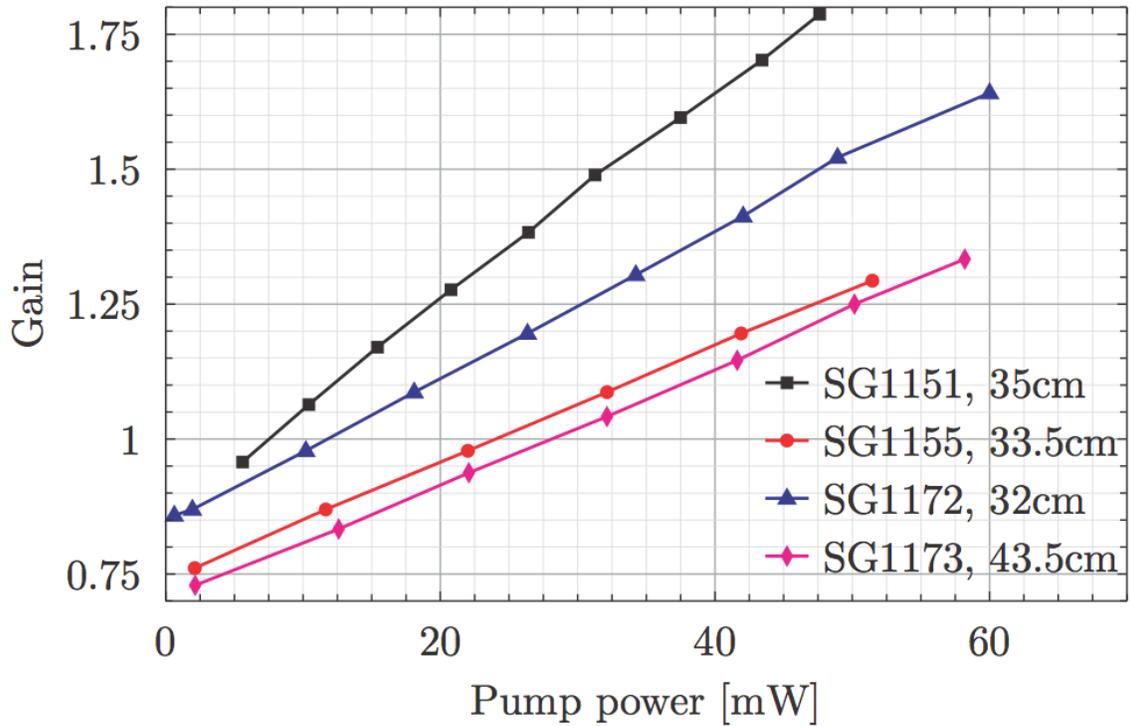

Figure 2. The gain of the selected fibers. The fibers are pumped at 1611 nm, the input signal has a power of 2.9 mW and a wavelength of 1945nm. The gain is defined as a ratio of the output power to the input power and includes the loss of splices with SMF28 at both ends.

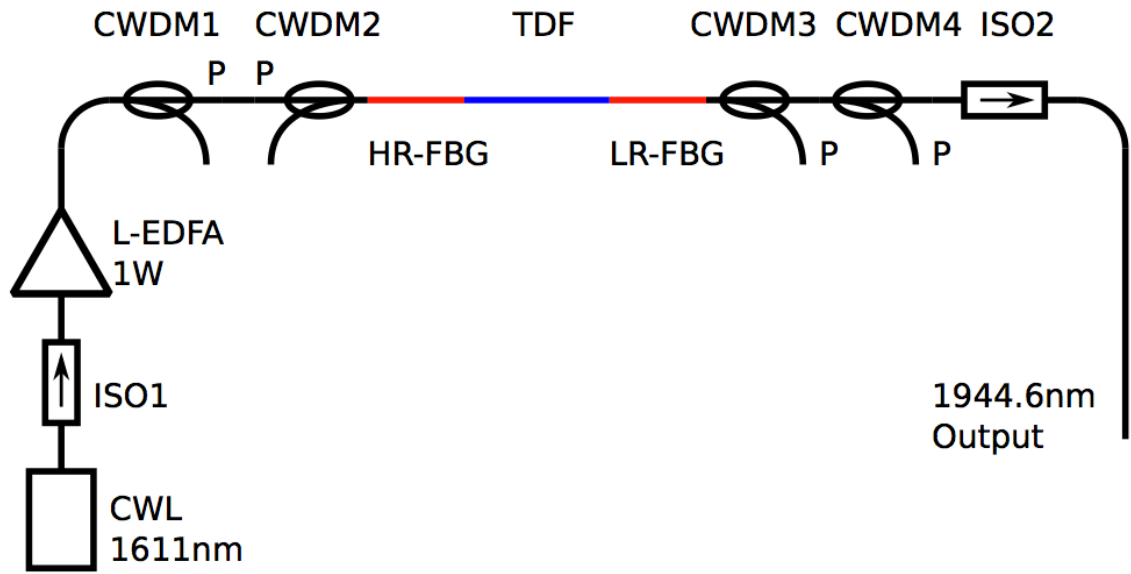

Figure 3. Set-up for testing the fiber in a laser experiment. ISO–isolators, HR-FBG–high reflectivity grating, LR-FBG–low reflectivity grating, CWDM–coarse wavelength division multiplexers.

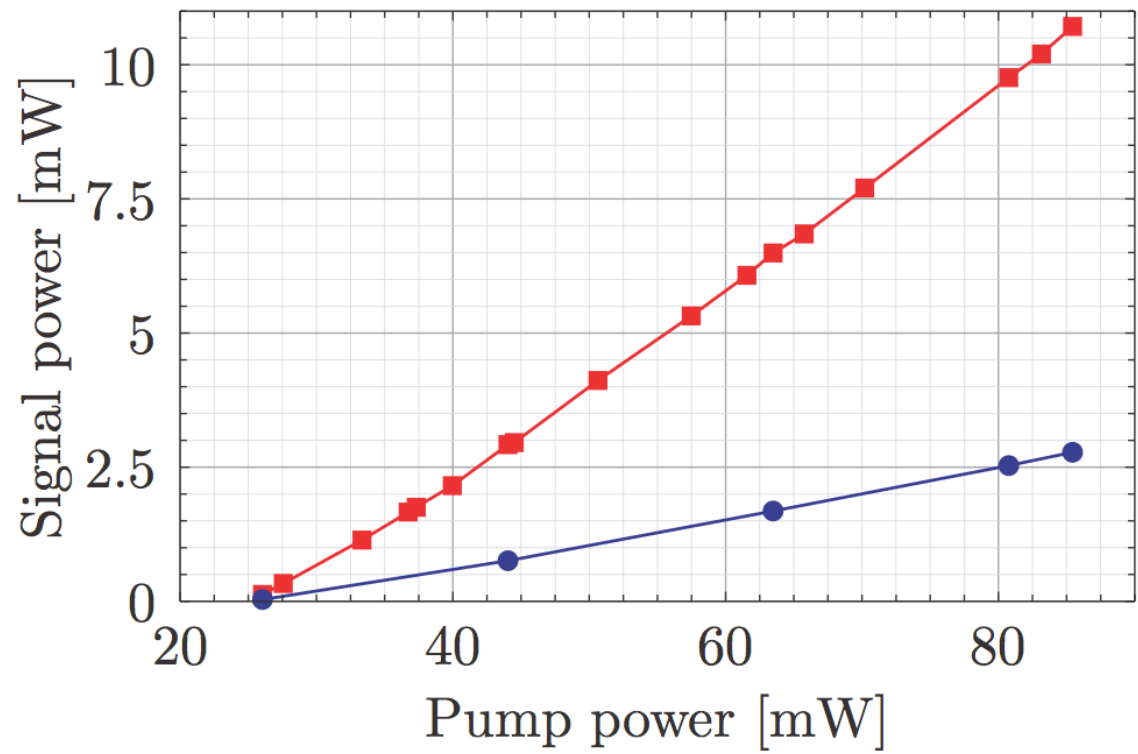

Figure 4. Results of testing SG1155 in the laser experiment. The signal power is measured as function of the pump power at the pump-end (circles) and at the opposite (output) end of the laser (squares).